\documentstyle[12pt,aaspp4]{article}

\begin{document}
{\hskip 10cm 
ApJ Lett. accepted}
\title{Prompt High Energy $\gamma$-Ray Emission From the Synchrotron Self-Compton Process
 in the Reverse 
Shocks of $\gamma$-Ray Bursts}

\author{X.Y. Wang,  Z.G. Dai  and T. Lu}
\affil{Department of Astronomy,
Nanjing University, Nanjing, P.R.China
}
\affil{Email:xywang@nju.edu.cn; daizigao@public1.ptt.js.cn; tlu@nju.edu.cn}

\begin{abstract}
In the standard scenario of the fireball model of gamma-ray bursts(GRBs),
the huge initial energy release produces a relativistic blast wave expanding
into the external medium and a reverse shock moving into and heating the fireball ejecta.
We calculate the high energy gamma-ray emission due to inverse Compton scattering of the 
synchrotron photons from relativistic electrons in the reverse shock.
Under the favorable values of the physical parameters of the GRBs and the interstellar
medium, our result shows that during the prompt phase,
this emission   dominates over the  component from
the forward shock at 
high energy gamma-ray bands. This mechanism can excellently account for 
the observations of the prompt high energy gamma-rays detected by EGRET, such as  from
GRB930131.

\end{abstract}


\section{Introduction}
Gamma-ray bursts (GRBs) are widely believed to be caused by the 
dissipation of the kinetic energy of a relativistically 
expanding fireball
with a Lorentz factor of $\eta\sim 10^2$--$10^3$
(see e.g. Piran 1999 for a review). 
After producing the prompt GRB, the shell impacts on the surrounding gas, driving an
ultra-relativistic shock into the ambient medium, which approaches a self-similar
behavior (Blandford \& McKee 1976) after a short transition phase. 
This forward shock continuously heats fresh gas
and accelerates electrons, producing a long-term afterglow.
The initial interaction between the ejecta with the surrounding gas also
leads to a reverse shock which propagates into and decelerates the ejecta.
This shock, heating up the  shell's matter and accelerating its electrons,
operates only once. 
Once the reverse shock crosses the ejecta, the ejecta expands and cools adiabatically.
Transition of the fireball to the self-similar behavior occurs
on a time scale comparable to the reverse shock crossing time of the ejecta.
Thus, emission from the fireball ejecta is suppressed after the transition  to the 
self-similar expansion. 

A prompt 9-th magnitude, optical flash has been recently
detected (Akerlof et al. 1999) accompanying  GRB990123. The most natural explanation
of this flash is synchrotron emission from a reverse shock propagating into the 
fireball ejecta after it interacts with the surrounding gas (Sari \& Piran 1999a;
M\'{e}sz\'{a}ros \& Rees 1999). The reverse shock, at its peak, contains energy which
is comparable to that of the GRB itself and the forward shock, but has much lower
temperature than that of the forward shock. So it radiates at considerably lower 
frequencies.
Since, during the prompt phase,
the number of heated electrons in the reverse shock is $\eta$($\sim 10^{2}-10^3$)
times larger than in the forward shock and  the seed photon source is stronger,
we expect a strong synchrotron self-Compton emission flux at  high energy gamma-ray  bands
 in the reverse shock.
Here we present an analytic calculation of this synchrotron self-Compton emission
 and consider its implications for the prompt high energy gamma-ray emission detected
from some GRBs  by the Energetic Gamma Ray Experiment Telescope
(EGRET) and Compton Telescope 
(COMPTEL)  on board Compton Gamma-Ray Observatory (CGRO). 
Early discussions on the synchrotron self-Compton process in the reverse shock include
the works of, e.g., M\'{e}sz\'{a}ros et al. (1993) and Sari et al. (1996), both of which,
however, are quite different from our  subject in many fundamental aspects.

\section{The synchrotron self-Compton emission from the reverse shock}

\subsection{Reverse shock model}

Transition of the fireball ejecta to the self-similar expansion occurs
on time scale $\Delta{t_{tran}}$ comparable to the longer of the two time scales
set by the initial conditions: the (observed) GRB duration $\Delta{t_{GRB}}$
and the (observed) deceleration time scale $t_{dec}$ of the fireball ejecta, at which
the self-similar Lorentz factor ($\Gamma$) equals the initial ejecta Lorentz factor
$\eta$, i.e. $\Delta{t_{tran}}=\max[\Delta{t_{GRB}}, t_{dec}]$
 (Sari \& Piran 1999b; Waxman \& Draine 2000).

The deceleration time 
$t_{dec}$ is in fact the observer time at which the heated interstellar medium (ISM)
energy is comparable to the initial energy $E$, i.e.
\begin{equation}
{t_{dec}}=\frac{r_{dec}}{2\eta^2{c}}(1+z)=10 {\rm sec} (\frac{1+z}{2}){E_{53}^{1/3}}
{n_0^{-1/3}}{\eta_{300}^{-8/3}},
\end{equation}
where 
\begin{equation}
r_{dec}=(\frac{E}{4\pi{\eta}^2{nm_{p}c^2}})^{1/3}
=3.7\times10^{16}{\rm cm}
{E_{53}^{1/3}}{n_0^{-1/3}}{\eta_{300}^{-2/3}}
\end{equation}
is the deceleration radius, $E=10^{53}E_{53}{\rm erg}$, $n=1n_0{\rm cm^{-3}}$,
$\eta=300\eta_{300}$, and $z$ is the redshift of the GRB source.
Since even  long duration GRBs have durations of order of $10{\rm sec}$, which
is comparable to ${t_{dec}}$, we therefore take $\Delta{t_{tran}}\sim{t_{dec}}$
for the following calculations. 

During the transition, the unshocked fireball shell propagates at the original expansion
Lorentz factor $\eta$ and the Lorentz factor of the plasma shocked by the reverse shock
in the rest frame of the unshocked ejecta is $\bar{\gamma}^{rs}\simeq{{\eta}/{\Gamma}}$.
According to Sari \& Piran (1995), the reverse shock Lorentz factor $\bar{\gamma}^{rs}$
depends on the parameter $\xi\equiv(l/\Delta)^{1/2}{\eta}^{-4/3}=2{E_{53}^{1/3}}
{n_0^{-1/3}}{\Delta_{10}^{-1/2}}{\eta_{300}^{-4/3}}$, where 
$l=(\frac{E}{nm_{p}c^2})^{1/3}$ is the Sedov length and $\Delta=c\Delta{t}_{GRB}$ is the
shell width.
If $\xi\gg1$, the reverse shock is Newtonian and $\bar{\gamma}^{rs}\sim{1}$, while
if $\xi\ll1$, the reverse shock is relativistic and $\bar{\gamma}^{rs}\gg1$.
Since the shell is likely to spread, which adjusts $\xi>1$ to $\xi\simeq{1}$
(Sari \& Piran 1995),
the reverse shock  we considered here will be 
always  mildly relativistic ($\bar{\gamma}^{rs}\sim1$)
 for typical parameter values of the fireball shock.

The distributions of the electrons in both the forward and reverse shocks are
assumed to be a power-law of index $p$ ($N(\gamma)\propto{\gamma}^{-p}$),
with the Lorentz factor of the random motion of a typical electron in the shell 
rest frame being $\gamma_{m}=\frac{m_p}{m_e}\xi_{e}{\gamma_{sh}}$,
where $\xi_{e}$ is the fraction of thermal energy carried by electrons
(Waxman 1997a,b ).
Here $\gamma_{sh}=\Gamma$ for the forward shock while $\gamma_{sh}=\bar{\gamma}^{rs}$
for the reverse shock.
Assuming that $\xi_B$ is the fraction of the thermal energy carried by the magnetic field,
 we have the peak frequency of the reverse shock emission
\begin{equation}
\nu_{m}^{rs}=\frac{1}{1+z}\frac{\Gamma({\gamma_m^{rs}})^2{eB'}}{2\pi{m_e}c}=
1.6\times10^{15}{\rm Hz}(\frac{\xi_{e}}{0.3})^2(\frac{\xi_B}{0.01})^{1/2}
\eta_{300}^2{n_0^{1/2}}(\frac{2}{1+z}),
\end{equation}
where $B'=12{\rm G}(\frac{\epsilon_B}{0.01})^{1/2}\eta_{300}n_0^{1/2}$,
is the magnetic field in the comving frame.
 Here $\xi_e$ and $\xi_B$ are the values relevant for
the reverse shock and have been assumed to be not different from the forward 
shock (Wang, Dai \& Lu 2000). The numerical values we have used are those
characteristic of the forward shock (Waxman 1997a; Freedman \& Waxman 2000;
Wijers \& Galama 1999; Granot et al. 1999). As this peak frequency 
we estimated  is slightly
higher than the optical band and it decreases with time after the deceleration
time $t_{dec}$, we expect that the optical light curve should posses a {\rm short}
initial rise phase, and then  decays with time,  which is consistent with
 the light curve  of the prompt optical flash from GRB990123.
 Waxman \& Draine (2000) have also derived the peak flux for the reverse shock
\begin{equation}
f_m^{rs}=1.5{\rm Jy}h_{65}^2(\frac{\sqrt{2}-1}{\sqrt{1+z}-1})^2(\frac{\xi_B}
{0.01})^{1/2}{\eta_{300}^{-1}}{n_0^{-1/4}}{E_{53}^{5/4}}(\frac{2}{1+z}\frac
{{t_{dec}}}{10{\rm sec}})^{-3/4},
\end{equation}
for a flat universe with zero cosmological constant and $H_0=65h_{65}
{\rm Km~s^{-1}{Mpc}^{-1}}$.
This flux is consistent with the observations of GRB990123 within a factor
of two, whose peak flux
is about $1{\rm Jy}$ (Akerlof et al. 1999).
The peak frequency of synchrotron emission from the forward
shock is larger than that of the reverse shock by a factor $\sim(\Gamma^2/\eta)^2$, and
is, at the deceleration time $t_{dec}$ of the ejecta,
\begin{equation}
\nu_m^{fs}=1.4\times10^{20}{\rm Hz}(\frac{\xi_e}{0.3})^2(\frac{\xi_B}{0.01})^{1/2}
\eta_{300}^4{n_0}^{1/2}(\frac{2}{1+z}),
\end{equation}
since at this time the self-similar Lorentz factor $\Gamma\simeq{\eta}$.

The cooling break frequencies , at which the cooling time scale of electrons 
equals that of the expansion dynamic time, are given by
\begin{equation}
\nu_c^{rs}=\frac{10^{17}\rm Hz}{(Y_{rs}+1)^2}(\frac{\xi_B}{0.01})^{-3/2}
\eta_{300}^{-4}{n_0^{-3/2}}(\frac{t_{dec}}{10\rm sec})^{-2}(\frac{1+z}{2})
\end{equation}
for the reverse shock, and
\begin{equation}
\nu_c^{fs}=\frac{10^{17}\rm Hz}{(Y_{fs}+1)^2}(\frac{\xi_B}{0.01})^{-3/2}
\eta_{300}^{-4}{n_0}^{-3/2}(\frac{t_{dec}}{10\rm sec})^{-2}(\frac{1+z}{2})
\end{equation}
for the forward shock, where $Y_{rs}$ and $Y_{fs}$ are the Compton parameters 
for the reverse shock and forward shock, respectively
(Panaitescu \& Kumar 2000).
The Compton parameter Y, expressing the cooling rate of electrons
due to inverse Compton effect, is defined as $Y=\frac{4}{3}\tau_{e}\int{\gamma}^2
\bar{N}_e(\gamma)d{\gamma}$, where$\bar{N}_e(\gamma)$ is the 
normalized electron distribution
and $\tau_{e}$ is the optical thickness to electron scattering.
According to Panaitescu \& Kumar (2000), 
$Y_{fs}=\frac{1}{2}(\sqrt{\frac{5}{6}\frac{\xi_e}{\xi_B}+1}-1)\sim2$,
since the forward shock is in the fast cooling regime, while
$Y_{rs}\sim(\gamma_m^{rs})^{p-1}(\gamma_c^{rs})^{3-p}{\tau_e^{rs}}\sim3$ 
for the reverse shock, where 
${\tau_e^{rs}}$ is the optical thickness of the shocked ejecta to electron scattering,
which will be calculated in the next subsection and
$\gamma_c$ is the Lorentz factor where the total energy emitted at  the time $t_{dec}$
is comparable to the electron's energy: $\gamma_c=\frac{6\pi{m_e}c(1+z)}
{\sigma_T\Gamma(B')^2{t_{dec}}}=
\frac{4500}{Y+1}(\frac{\epsilon_B}{0.01})^{-1}
{\eta_{300}^{-3}}(\frac{t_{dec}}{10\rm sec})^{-1}(\frac{1+z}{2})n_0^{-1}$.
The synchrotron spectrum of the forward shock is thus described as
$f_{\nu}^{fs}=f_{m}^{fs}(\nu/{\nu_c^{fs}})^{-1/2}$ for $\nu_m^{fs}>\nu>\nu_c^{fs}$
and $f_{\nu}^{fs}=f_m^{fs}(\nu_m^{fs}/\nu_c^{fs})^{-1/2}(\nu/\nu_m^{fs})^{-p/2}$
for $\nu>\nu_m^{fs}$; while for the reverse shock,
$f_{\nu}^{rs}=f_m^{rs}(\nu/\nu_m^{rs})^{-\frac{p-1}{2}}$ for $\nu_c^{rs}>\nu>\nu_m^{rs}$
and $f_{\nu}^{rs}=f_m^{rs}(\nu_c^{rs}/\nu_m^{rs})^{-\frac{p-1}{2}}(\nu/\nu_c^{rs})^
{-\frac{p}{2}}$ for $\nu>\nu_c^{rs}$.

\subsection{The synchrotron self-Compton emission from the reverse shock}
The synchrotron self-Compton emission from
the reverse shock can be calculated by using the 
above equations for the synchrotron spectrum and the Thomson optical depth of the 
shocked fireball material,
which is 
\begin{equation}
\tau_e^{rs}=\frac{\sigma_T{N}}{4\pi{r_{dec}^2}}
=8.4\times10^{-6}{E_{53}^{1/3}}
{n_0^{2/3}}{\eta_{300}^{1/3}}(\frac{t_{obs}}{t_{dec}})^{-1/2},
\end{equation}
where $N$ is the total number of electrons in the fireball and $\sigma_T $ is the 
Thomson scattering cross section.
This inverse Compton luminosity for the reverse shock is larger than the synchrotron
luminosity by a factor $L_{IC}/L_{sy}\sim{Y_{rs}}\sim3$ ,  dominating  the 
synchrotron cooling  rate of the electrons, and radiates at much higher energy band than
the synchrotron emission.
We need consider only the first-order inverse Compton and neglect the higher order
inverse Compton process, because a once-scattered synchrotron photon has energy
of order $(h\nu_{min})_{com}{\gamma_e^3}>m_{e}c^2$ in the rest frame of the 
second scattering electron with Lorentz factor being $\gamma_e$ 
and thus the Thomson limit no longer applies.
Since the reverse shock is in the slow cooling regime,
the up-scattered spectrum peaks at 
\begin{equation}
\nu_m^{rs,IC}=2(\gamma_m^{rs})^2{\nu_m^{rs}}=1.0\times10^{21}{\rm Hz}(\frac{\xi_e}
{0.3})^4(\frac{\xi_B}{0.01})^{1/2}{\eta_{300}^2}{n_0^{1/2}}   
(\frac{\bar\gamma_{rs}}{1})^2(\frac{2.0}{1+z}),
\end{equation}
and the cooling break frequency, at the deceleration time $t_{dec}$, is
\begin{equation}
\nu_c^{rs,IC}=2(\gamma_c^{rs})^2\nu_c^{rs}=2.1\times10^{22}{\rm Hz}
(\frac{\xi_B}{0.01})^{-7/2}{\eta_{300}^{2/3}}{n_0^{-13/6}}{E_{53}^{-4/3}}(\frac{2}{1+z}),
\end{equation}
where we have substituted the expression of $t_{dec}$ (Eq.1)into it.
At this deceleration time of the 
ejecta, the peak flux of the inverse Compton component  is
\begin{equation}
f_{max}^{rs,IC}=\tau_e^{rs}f_{m}^{rs}=2.6\times10^{-8}{\rm erg~cm^{-2}s^{-1}MeV^{-1}}
E_{53}^{4/3}{n_0^{7/6}}{\eta_{300}^{4/3}}(\frac{\xi_B}{0.01})^{1/2}h_{65}^2
(\frac{\sqrt{2}-1}{\sqrt{1+z}-1})^2.
\end{equation}
For the simplicity of the analysis, we use the broken-power law approximation
to the inverse Compton component,
though the presence of logarithmic terms makes some corrections to the spectrum
for $\nu>\nu_m^{rs,IC}$ (Sari \& Esin 2000), i.e.
$f_{\nu}^{rs,IC}=f_m^{rs,IC}(\nu/\nu_m^{rs,IC})^{-\frac{p-1}{2}}$ for 
$\nu_c^{rs,IC}>\nu>\nu_m^{rs,IC}$
and $f_{\nu}^{rs,IC}=f_m^{rs,IC}(\nu_c^{rs,IC}/\nu_m^{rs,IC})^{-\frac{p-1}{2}}
(\nu/\nu_c^{rs,IC})^{-\frac{p}{2}}$ for $\nu>\nu_c^{rs,IC}$.
For the sake of comparison, we give the flux of the inverse Compton component
at three representative frequencies:
\begin{eqnarray}
f^{rs,IC}(\varepsilon=5{\rm MeV})=2.6\times10^{-8}{\rm erg~cm^{-2}s^{-1}MeV^{-1}}
E_{53}^{4/3};  \nonumber \\
f^{rs,IC}(\varepsilon=100{\rm MeV})=2.7\times10^{-9}{\rm erg~cm^{-2}s^{-1}MeV^{-1}}
E_{53}^{4/3};  \nonumber \\
f^{rs,IC}(\varepsilon=1{\rm GeV})=1.5\times10^{-10}{\rm erg~cm^{-2}s^{-1}MeV^{-1}}
E_{53}^{2/3},
\end{eqnarray}
where we have only retained the relations with the shock energy $E_{53}$ and
replaced other shock parameters and the redshift with the typical values.

After the reverse shock has passed through the ejecta, the ejecta cools adiabatically.
Sari \& Piran (1999a) assumed that the ejecta follows the Blandford-McKee (1976) 
self-similar solution, in which a given fluid element evolves with a bulk Lorentz
factor of $\gamma\sim{R}^{-7/2}$, and successfully explained the decaying light curve
of the optical flash and the radio flare behavior of  GRB990123 (Kulkarni et al. 1999).
Below we will give the decaying light curve of the synchrotron self-Compton
component of the reverse shock after it has passed through the ejecta ($t_{obs}>t_{dec}$).
Since $f_m^{rs}\propto{t_{obs}^{-47/48}}\sim{t_{obs}^{-1}}$ and
 $\tau_e^{rs}\propto{t_{obs}^{-1/2}}$,
the peak flux of the inverse Compton spectral component $f_m^{rs,IC}=f_m^{rs}\tau_e^{rs}
\propto{t_{obs}^{-3/2}}$.
If the observed frequency locates between the two break frequencies ($\nu_c^{rs,IC}
<\nu<\nu_m^{rs,IC}$), then $f_\nu^{rs,IC}=f_m^{rs,IC}(\frac{\nu}{\nu_m^{rs,IC}})^{-(p-1)/2}$.
According to Sari \& Piran (1999a), $\gamma_m\propto{t_{obs}^{-13/48}}$ and
$\nu_m^{rs}\propto{t_{obs}^{-73/48}}$, so $\nu_m^{rs,IC}=2\gamma_m^2{\nu_m^{rs}}
\propto{t_{obs}^{-33/16}}\sim{t_{obs}^{-2}}$, thus $f_\nu^{rs,IC}\propto{t_{obs}^{-1/2-p}}\propto{t_{obs}^{-3}}$
for $p=2.5$.
On the other hand, if the observed frequency is above $\nu_c^{rs,IC}$, the flux drops
exponentially with time since all the electrons above the corresponding energy cool and no 
fresh electrons are accelerated once the reverse shock has crossed the ejecta shell.
Therefore, in general, we expect to see a rapidly decaying high energy flux from the 
reverse shock, which constitutes a unique characteristic distinguished 
from other models
(see the next subsection) suggested for the observed high energy gamma-ray 
emission 
from some GRBs.
Measurements of the time dependence of the high energy gamma-ray 
flux  with the planned 
Gamma-ray Large Area Space Telescope (GLAST) mission will test this synchrotron
self-Compton scenario. 

\subsection{Detectablity of the prompt high energy  gamma-rays and the case
of GRB930131} 

EGRET has detected {\em prompt} emission above $30{\rm MeV}$ from seven  bright GRBs
triggered by BATSE (Catelli et al. 1998), among which $1.2{\rm GeV}$ and $3.4{\rm GeV}$
photons have been detected from GRB930131 (Sommer et al. 1994; Ryan et al. 1994)
and GRB940217 (Hurley et al. 1994), respectively.
GRB940217 also displays delayed high energy gamma-ray emission, about 90 minutes after the initial
trigger. 
There are a handful models suggested to explain the delayed and prompt GeV emission.
For examples,
Katz (1994) suggested that the impact of the fireballs with dense clouds ($n\geq{10
^9{\rm cm^{-3}}}$) could yield high energy gamma-ray emission via $\pi^0$ decay;
Waxman \& Coppi (1996) proposed that the cascading of ultra high energy ($\geq10^{19}
{\rm eV}$) cosmic rays (accelerated in cosmological GRBs)  off infra-red and
cosmic microwave background fields can produce delayed GeV-TeV emission due
to the deflections by the intergalactic magnetic field;
 Dermer et al. (2000) argued that the synchrotron
self-compton emission from the forward shock expanding into a 
rather dense circumburst medium  ($n\sim100{\rm cm^{-3}}$)
 may be responsible for 
the prompt and delayed
high energy gamma-ray emission; Vietri (1997) and Totani (1998) suggested that the synchrotron
emission of the protons may be responsible for the GeV-TeV emission. 
Here we suggest that the synchrotron self-Compton emission from 
the reverse shock could explain both the flux level and the spectrum of the 
high energy gamma-rays detected by EGERT. As an example, we discuss the case of GRB930131.
The high energy gamma-ray emission from GRB930131 extends for about $25{\rm s}$.
Totally sixteen gamma rays above $30{\rm MeV}$ were detected, including two with energy
about $1{\rm GeV}$, and the power-law fit of their  photon spectrum is given by
$\frac{dn}{d\varepsilon}\sim7.4\times10^{-6}{\rm photons~(cm~s~MeV)^{-1}}
(\varepsilon/147{\rm MeV})^{-2.07\pm0.36}$ (Sommer et al. 1994).
In the last subsection, we have found that the inverse Compton emission intensity 
from the reverse shock is $f_{\nu}^{rs,IC}\sim
2.7\times10^{-9}{\rm erg~cm^{-2}s^{-1}MeV^{-1}}(\varepsilon/{\rm 100MeV})^{-0.75}
E_{53}^{4/3}$ for $\varepsilon<h{\nu_c^{rs,IC}}$ and
$f_{\nu}^{rs,IC}\sim
2.7\times10^{-9}{\rm erg~cm^{-2}s^{-1}MeV^{-1}}(\varepsilon/{\rm 100~MeV})^{-1.25}
E_{53}^{2/3}$ for $\varepsilon>h{\nu_c^{rs,IC}}$
with $p=2.5$, 
that is, the differential photon flux is $\frac{dn}{d\varepsilon}\sim
6.0\times10^{-6}{\rm photons~(cm~s~MeV)^{-1}}(\varepsilon/147{\rm MeV})^{-2.25}
E_{53}^{2/3}$. Thus, if the fireball shock energy $E\sim1.5\times10^{53}{\rm erg}$
and other parameters such as $\xi_e$, $\xi_B$, $\eta$, $z$ and the number density
$n$
of the surrounding medium take the above representative values, then both the flux level
and the spectrum agree well with the observations.
However, since this emission process lasts only a time comparable to the crossing time
of the reverse shock, it cannot explain the delayed high energy gamma-ray emission, as observed
from GRB940217, which may, for example, be due to deflections of ultra high
energy cosmic rays accelerated in the GRB fireballs by the 
intergalactic field (see Waxman \& Coppi 1996).

\subsection{Comparision with the high energy gamma-ray emission flux from the 
forward shock}
For completeness, we also compute the high energy gamma-ray emission flux from the synchrotron
and self-Compton process in the forward shock,
 and compare them with that from the reverse shock.
The peak flux of the forward shock is 
\begin{equation}
f_m^{fs}=8{\rm mJy}{h_{65}^2}(\frac{\sqrt{2}-1}{\sqrt{1+z}-1})^2(\frac{\xi_B}
{0.01})^{1/2}E_{53}{n_0^{1/2}}
\end{equation}
(Waxman 1997b; Wijers \& Galama 1999), 
and the two break frequencies $\nu_m^{fs}$ and $\nu_c^{fs}$
are given by Eq.(5) and Eq.(7). Please note that here the flux peaks at $\nu_c^{fs}$.     
For typical parameter values of GRBs, during the prompt phase the forward shock is in the 
fast cooling regime, and we get the high energy flux at those three representative
frequencies:
\begin{eqnarray}
f^{fs}(\varepsilon=5{\rm MeV})=1.2\times10^{-8}{\rm erg~cm^{-2}s^{-1}MeV^{-1}}
E_{53};  \nonumber \\
f^{fs}(\varepsilon=100{\rm MeV})=2.5\times10^{-10}{\rm erg~cm^{-2}s^{-1}MeV^{-1}}
E_{53};  \nonumber \\
f^{fs}(\varepsilon=1{\rm GeV})=1.4\times10^{-11}{\rm erg~cm^{-2}s^{-1}MeV^{-1}}
E_{53}.
\end{eqnarray}
We can see that the flux at $\sim 5{\rm MeV}$ is comparable to the inverse Compton
component of the reverse shock, while at  higher energies ($>100{\rm MeV}$)
it is about one order of magnitude  lower.
As in the reverse shock, we also need consider the synchrotron self-Compton emission process
in the forward shock, since the typical energy of the synchrotron photon in the 
rest frame of the scattering electrons is of the order of 
$(h\nu_c^{fs})_{com}{\gamma_e}{\ll}m_e{c^2}$, which is still in the Thomson limit range.
At the deceleration time ($t_{obs}\simeq{t_{dec}}$),
the characteristic break frequencies of the inverse Component of the forward shock
are 
\begin{equation}
\nu_m^{fs,IC}=2(\gamma_m^{fs})^2\nu_m^{fs}=8.0\times10^{30}{\rm Hz}
(\frac{\xi_e}{0.3})^4(\frac{\xi_B}{0.01})^{1/2}{\eta_{300}^6}{n_0}^{1/2}
(\frac{2.0}{1+z})
\end{equation}
and
\begin{equation}
\nu_c^{fs,IC}=2(\gamma_c^{fs})^2\nu_c^{fs}=5.0\times10^{22}{\rm Hz}
(\frac{\xi_B}{0.01})^{-7/2}{\eta_{300}^{2/3}}{n_0^{-13/6}}{E_{53}^{-4/3}}(\frac{2}{1+z}).
\end{equation}
The peak flux of this inverse Compton component is
\begin{equation}
f_m^{fs,IC}=\tau_e^{fs}f_m^{fs}\sim{\sigma_T}nr{f_m^{fs}}=3\times10^{-13}
{\rm erg~cm^{-2}s^{-1}MeV^{-1}}(\frac{r}{r_{dec}})E_{53}(\frac{\xi_B}{0.01})^{1/2}
n_0^{3/2}.
\end{equation}
Therefore the inverse Compton flux from the forward shock 
at $\varepsilon\sim{\rm 250MeV}$ band is $\sim3\times10^{-13}
{\rm erg~cm^{-2}s^{-1}MeV^{-1}}$, much lower than that from the reverse shock.
They are comparable only at $\varepsilon\sim{\rm 10TeV}$ band.
So, we conclude that for the typical parameter values of the shock and the 
surrounding medium, the synchrotron self-Compton emission from
the reverse shock dominates over the
synchrotron and synchrotron self-Compton emissions 
from the forward shock at  high energy gamma-ray bands.
Our result is different from that of
Dermer et al. (2000), who argue that the synchrotron
self-Compton emission from the forward shock may be responsible for 
the prompt and delayed
high energy gamma-ray emission.
 The key point of the difference is that 
they considered a rather dense circumburst medium with number density 
 $n\sim100{\rm cm^{-3}}$, while
we consider a typical interstellar medium  with $n\sim1{\rm cm^{-3}}$.

\section{ Summary and Discussion}
The detection of the prompt optical flash and the radio flare from GRB990123
has given positive evidence for the reverse shock model. In this paper,
we calculated analytically the synchrotron self-Compton emission from the 
reverse shock. Assumed favorable parameter values for the fireball shock
and the density of the interstellar medium to be 
$\xi_e=0.3$, $\xi_B=0.01$, $E_{53}=1$, 
$p=2.5$, $\eta=300$ and $n=1{\rm cm^{-3}}$,
the result shows that during the prompt phase, this emission dominates over
the emission due to other processes of the shocked electrons in both the forward
 and reverse external
shocks at high energy gamma-ray  bands. We suggest that this process provides a new explanation
for the {\em prompt} high energy gamma rays detected by EGRET from some bright
bursts, such as GRB930131, GRB910503 and GRB940217 etc.
As an example, we compared our calculated result with the observations of 
GRB930131 and found that both the flux level and spectrum agree quite well,
 provided the initial fireball shock energy is a few times  $10^{53}{\rm erg}$,
which is consistent with the fact that all the bursts detected by EGRET
are bright bursts in the BATSE range.
We also derived the decaying light curves of the inverse Compton component
and found that it decays quite rapidly, regardless
whether the observed band locates above or below the cooling break frequency
of the inverse Compton component. This unique feature can act as a test of 
our model if in the future GLAST is  capable of describing the time evolution
of the high energy gamma-ray flux. 
Because once the reverse shock has crossed that ejecta, 
all electrons cool by adiabatic expansion and no fresh electrons
are accelerated, our model can not explain the delayed ${\rm GeV}$ emission,
as observed from GRB940217.

 GRBs are generally believed to be produced by the internal collisions between
the relativistic shells at a radius $r\sim10^{12}-10^{13}{\rm cm}$, which is 
much smaller than the distance at which the external shock takes place.
Till now, the mechanism of  prompt GRB emission is not as well-understood as that
of afterglows and we do not know whether the usual GRBs and high
energy gamma-rays have a common origin . If we here take a naive assumption that 
the internal shock emission extends from BATSE band to GeV 
bands with $f_{\nu}\propto{\nu}^{-p/2}$, the simple extrapolation gives a flux
of $f_{\nu}(\varepsilon=100{\rm MeV})\sim1.0\times10^{-8}{\rm erg~cm^{-2}s^{-1}MeV^{-1}}$ for typical GRB
fluence of $F_{\gamma}(\rm 20~KeV-2~MeV)\sim10^{-5}{\rm erg~cm^{-2}}$.
This flux at $\varepsilon=100`{\rm MeV}$ is comparable to that from
the reverse shock. However, this simple extrapolation may be incorrect if there is a sharp
cutoff at some high energy band, as shown in the model 
of Ghisellini \& Celotti (1999), in which the prompt GRB emission is due to quasi-thermal
comptonization process in internal shocks and  ends with an  
exponential cutoff at energies higher than a few  MeV. 

In all our calculation, we considered a fireball shock expanding into 
a uniform interstellar medium  
with constant density of $n\sim1{\rm cm^{-3}}$.
The observations of some afterglows have shown that some GRBs may occur in a stellar
wind environment (Dai \& Lu 1998; Chevalier \& Li 1999), 
implying the massive star origin of these bursts (Woosley 1993; Paczy{\'n}ski 1998).
As having been pointed out in Chevalier \& Li (2000) and Dai \& Lu (2000), 
the peak flux of the 
reverse shock formed by
the initial interaction between the fireball ejecta with the stellar wind medium
is significantly lower than that in the case of the interstellar medium
for similar shock parameters .
For this reason, we conjecture that the high energy gamma-ray  emission in this kind of reverse
shock  is too weak to be detected by high energy gamma-ray  detectors.

X.Y.Wang would like to thank Dr D. M. Wei and Z. Li for valuable discussions. 
This work was supported by the National Natural Science Foundation
of China under grants 19773007, 19973003 and 19825109 and the Foundation 
of the Ministry of Education of China.

\end{document}